 \documentclass[structabstract]{aa}  
 \usepackage{color} 
                    
\usepackage{graphicx}
\usepackage[]{natbib}
\usepackage{txfonts}

\newcommand{\be}{\begin{equation}}
\newcommand{\ee}{\end{equation}}
\newcommand{\beq}{\begin{eqnarray}}
\newcommand{\eeq}{\end{eqnarray}}
\newcommand\deriv[2]{ \frac{\mathrm d #1}{\mathrm d #2}}
\newcommand{\tang}{ \mathbf{ T} }
\newcommand\diff{ {\mathrm d} }

\begin{document}
\title{The writhe of helical structures in the solar corona}

\author{T. T\"or\"ok\inst{1,2}
        \and
        M. A. Berger \inst{2,3}
        \and
        B. Kliem    \inst{2,4,5}}

 \institute{LESIA, Observatoire de Paris, CNRS, UPMC, Universit\'e Paris Diderot,
            5 place Jules Janssen, 92190 Meudon, France\\
            \email{tibor.torok@obspm.fr}
            \and
            University College London, Mullard Space Science Laboratory,
            Holmbury St.~Mary, Dorking, Surrey, RH5 6NT, UK            
                        \and
            University of Exeter, SECAM, Exeter, EX4 4QE, UK
            \and
            Universit{\"a}t Potsdam, Institut f{\"u}r Physik und
            Astronomie, 14482 Potsdam, Germany
            \and
            Naval Research Laboratory, Space Science Division, Washington, DC 20375, USA}

\date{Received 1 November 2009; accepted ...}

\abstract
{
Helicity is a fundamental property of magnetic fields, conserved in ideal MHD. 
In flux rope topology, it consists of twist and writhe helicity. 
Despite the common occurrence of helical structures in the solar atmosphere, 
little is known about how their shape relates to the writhe, 
which fraction of helicity is contained in writhe, and 
how much helicity is exchanged between twist and writhe when they erupt. 
}
{
Here we perform a quantitative investigation of these questions relevant for 
coronal flux ropes. 
}
{
The decomposition of the writhe of a curve into local and nonlocal components
greatly facilitates its computation. We use it to study the relation 
between writhe and projected S shape of helical curves and to measure writhe and twist 
in numerical simulations of flux rope instabilities. 
The results are discussed with regard to filament eruptions and coronal mass 
ejections (CMEs). 
}
{
(1) We demonstrate that the relation between writhe and projected S shape is 
\emph{not} unique in principle, but that the ambiguity does not affect
low-lying structures, thus supporting the established empirical rule which 
associates stable forward (reverse) S shaped structures low in the corona 
with positive (negative) helicity.
(2) Kink-unstable erupting flux ropes are found to transform a far smaller 
fraction of their twist helicity into writhe helicity than often assumed. 
(3) Confined flux rope eruptions tend to show stronger writhe at low heights 
than ejective eruptions (CMEs). This argues against suggestions that the 
writhing facilitates the rise of the rope through the overlying field. 
(4) Erupting filaments which are S shaped already before the eruption and 
keep the sign of their axis writhe (which is expected if field of one 
chirality dominates the source volume of the eruption), must reverse their 
S shape in the course of the rise. Implications for the occurrence of the 
helical kink instability in such events are discussed. 
(5) The writhe of rising loops can easily be estimated from the angle of 
rotation about the direction of ascent, once the apex height exceeds the 
footpoint separation significantly. 
}
{
Writhe can straightforwardly be computed for numerical data and can often
be estimated from observations. It is useful in interpreting S shaped
coronal structures and in constraining models of eruptions.
}


\maketitle

\section {Introduction}
\label{sec:int}
Observations of the solar corona display a variety of structures that appear 
S shaped when viewed in projection on the disk, as for example filaments
with curved ends, soft X-ray sigmoids, and magnetic loops
that connect different active regions.
The presence of an S shape is regarded to be evidence for current-carrying 
twisted or sheared magnetic fields which possess magnetic helicity. 
The link to helicity is underlined by the fact that the orientation of the S 
shows a hemispheric preference, as other indicators of magnetic chirality 
(the helicity sign) do \cite[e.g.,][]{rus96, Zirker&al1997}. Coronal currents 
store the energy required to power eruptions \citep{for00}. Indeed, it has been 
shown that active regions exhibiting a sigmoidal morphology are more likely 
to erupt than non-sigmoidal ones \citep[][]{can99}. 

Twisted or sheared coronal fields carrying nearly force-free volume currents, 
especially magnetic flux ropes, are central in models of filaments and prominences 
\citep[e.g.,][]{aul98,Bobra&al2008}, of sigmoids 
\citep[e.g.,][]{rus96,tit99,low03,gib04,kli04}, 
and of eruptions
\citep[e.g.,][]{Forbes&Isenberg1991,ant99,ama00,fan03,toe05,Yeates&Mackay2009}. 
The amount of twist or shear in them 
can be quantified by the magnetic helicity. For a magnetic flux rope, 
the helicity is proportional to the sum of its twist and writhe. The twist 
measures how much the field lines wind about the magnetic axis of the rope, 
whereas the writhe quantifies the helical deformation of the axis itself.

The orientation of the S in helical structures on the Sun is highly 
correlated with the prevailing sign of magnetic helicity in their source volume. 
For example, \cite{Rust&Martin1994} found a one-to-one correlation between
the chirality of sunspot whirls and the orientation (sinistral
vs.\ dextral) of the axial field of filaments that spiral into their
periphery. Here, an apparent, or true, clockwise (counterclockwise)
rotation of the sunspot, associated with positive (negative) helicity of
its field, corresponds to forward (reverse) S shape of the filament end.
\cite{pev97} found that forward (reverse) S-shaped sigmoids in active regions 
were formed in predominantly positive 
(negative) helicity regions for $\approx$\,90\,\% of the cases investigated. 
In the remaining cases the helicity sign of the host active region was 
ambiguous, which is consistent with the possibility of a universal relation
between the orientation of the S in such structures and their chirality. In a study 
of rotating erupting filaments, \cite{gre07} found a one-to-one relationship between 
the sign of dominant helicity, the orientation of associated sigmoids,
and the orientation of the rising filament's developing S shape
(but see \citeauthor{Muglach&al2009} \citeyear{Muglach&al2009} for
occasional exceptions in the latter association).

In addition to stable structures, erupting filaments and prominences can
also exhibit helical deformations. These eruptions are seen as rising
loops in the corona, which in most cases become the core of a coronal
mass ejection (CME), or else fall back in a confined eruption. Their
change of shape can be described as a writhing of the filament axis out
of the plane defined by the filament's foot points and apex, or equivalently, 
as a rotational motion of the filament axis in the upper part of the loop about 
the direction of ascent. The total writhe varies widely from event to event, 
and is often acquired to a large part already in the low and middle corona. 
Some very clear cases of such writhing have been described by 
\cite{ji03}, \cite{rom03}, \cite{wil05}, and \cite{Zhou&al2006}. These events 
include a wide range of dynamical behaviour, from a confined eruption to the 
fastest CME so far recorded.
The writhing in the course of eruptions has attracted interest for three main 
reasons.

\begin{figure}[t]
\centering
\includegraphics[width=1.0\linewidth]{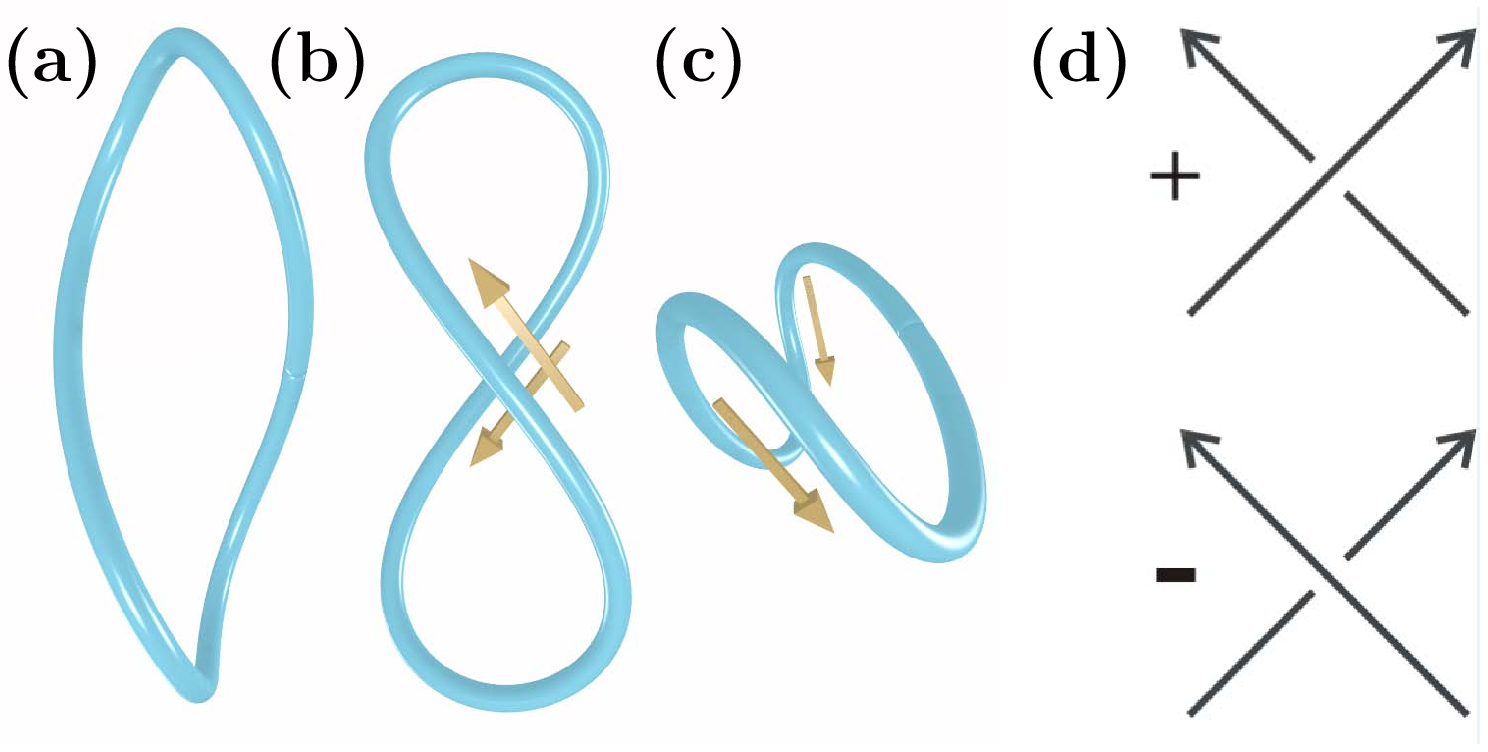}
\caption
{
A closed flux tube as seen from three different angles. The number of 
apparent crossings one sees depends on the viewing angle. Crossings can 
be called left--handed (negative) or right--handed (positive), depending 
on the relative orientations of the upper and lower field lines. The 
diagrams shown in {\bf (d)} give the two possible cases (up to rotation). 
The tube shown has an average crossing number, or \emph{writhe}, of 
$W = 0.566$. Most viewing angles display a positive crossing, as in 
{\bf (b)}. However, some angles do not show any crossings, as in {\bf (a)}, 
and there is a small set of angles which display a negative crossing, as 
in {\bf (c)}. (The latter picture of the tube is obtained by viewing it 
from the top.)
}
\label{fig:crossings}
\end{figure}

First, it indicates that the erupting field has the 
\emph{magnetic structure} of a freely moving flux rope, line tied only 
at its ends, at the onset of the helical deformation \cite[e.g.,][]{rus03, gre07}. 

Second, the observed deformations correspond exactly to the evolution of the 
helical kink instability of the current channel in the core of a flux rope 
(hereafter KI) and have, therefore, been taken as strong indication of this 
instability's occurrence \citep{sak76, rom03, rus05, toe05, wil05, Zhou&al2006, 
gil07}. The KI transforms some of the twist of the field lines about the magnetic
axis of the rope into writhe of the axis.
In doing so, it lowers the magnetic energy by reducing the tension in the 
twisted field. Since this is an ideal MHD instability and the corona is a 
nearly perfectly conducting medium, the conversion of twist into writhe is
constrained by the approximate conservation of the magnetic helicity
contained in the rope \citep{Berger1984}.
It has been suggested that the writhing of the flux rope's upper part into the 
direction of the overlying field is energetically favourable for its passage 
through the overlying field to become a CME \citep{Sturrock&al2001, fan05}. 
Thus, the deformations yield hints on the \emph{physical processes of 
CME initiation}.

Third, the writhing is the major factor in determining the final magnetic orientation 
of the CME (apart from influences during the interplanetary propagation), which, 
in turn, is one of the critical parameters that control the strength of the 
interaction if the CME hits the terrestrial magnetosphere---its 
\emph{geoeffectiveness}. The final orientation can differ largely from the erupting 
structure's initial orientation: rotations as large as $160^\circ$ have been 
reported \citep[albeit the estimated angles are very approximate, see]
[and references therein]{dem08, yur08}.

\begin{figure}[t]
\sidecaption
\centering
\includegraphics[width=0.61\columnwidth]{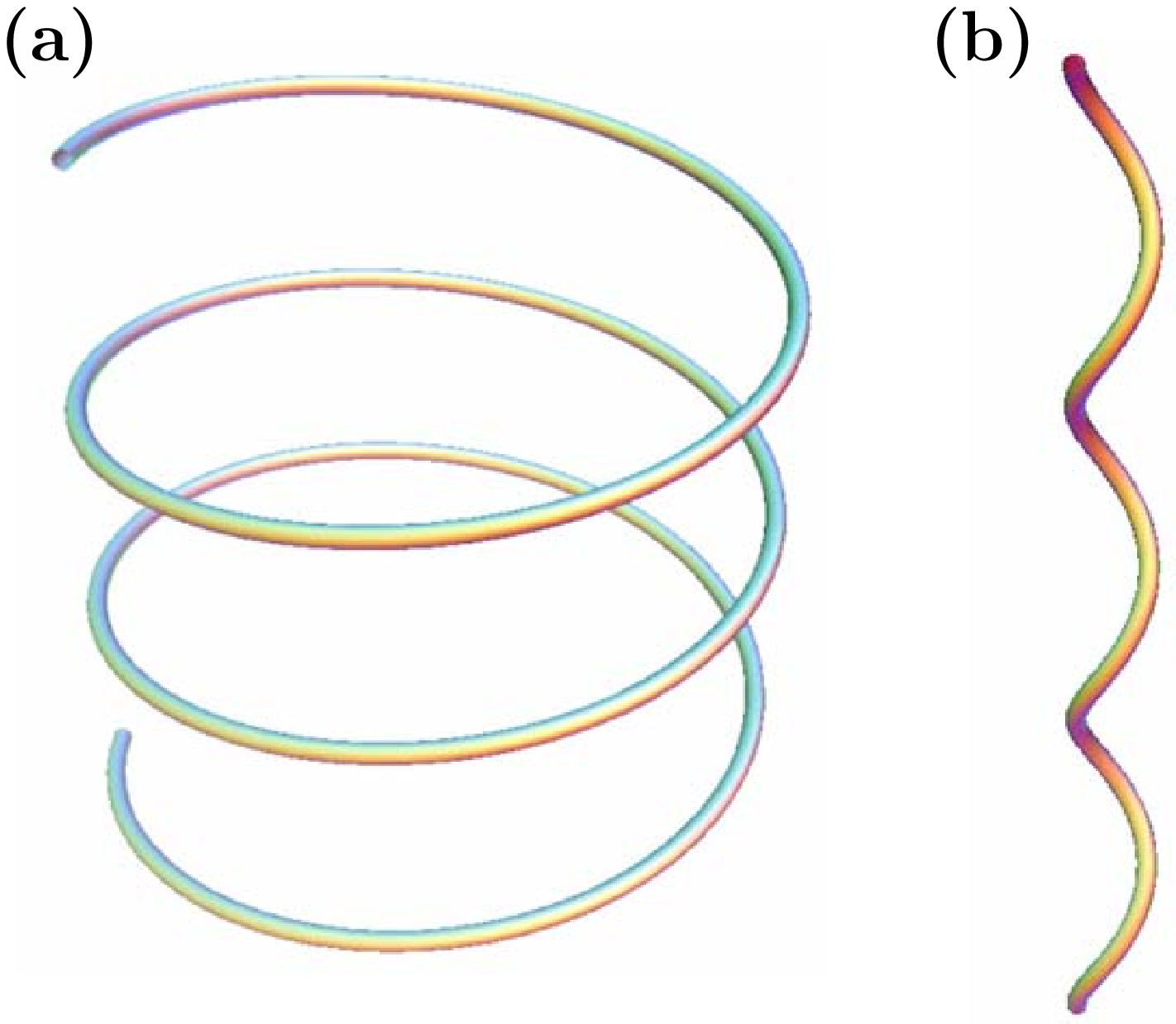}
\caption
{
The writhe of a coil depends on its aspect ratio (height/width) as well as 
how many turns it has. Both coils shown have 3 turns, but the fat coil in 
{\bf (a)} has a writhe $W = 2.68$, while the thin coil in {\bf (b)} has a writhe 
$W = 0.46$.
}
\label{fig:coils}
\end{figure}

Despite the common occurrence of both stable and dynamic helical structures on 
the Sun, measurements of the axis writhe of such structures have not yet been 
undertaken. So far, only estimates based on qualitative considerations have 
been reported \citep[e.g.,][]{vrs93, rus03}. This is mainly due to the fact 
that the observations have been limited to a 2D projection of intrinsically 3D 
structures onto the plane of the sky. With the advent of the \textsl{STEREO} 
mission, true 3D reconstructions of coronal structures are possible, 
as long as the angular separation between the twin satellites
or between one of them and the Earth stays within certain
intervals. Even for numerical simulations 
which provide the 3D magnetic field of evolving unstable flux ropes, the 
temporal evolution of twist and writhe has only extremely rarely been 
quantified \citep{lin98}. 
This is mainly due to the fact that the calculation of these quantities 
based on their general definitions is a relatively complicated task. 
The computation of the writhe is greatly facilitated by its decomposition into 
local and nonlocal components in \citet{ber06}. Here we make use of these 
expressions.

Our aim in this paper is to describe and discuss possible applications of 
writhe measurements for both stable and erupting objects in the solar corona. 
To this end, we first review, in Sect.~\ref{sec:def}, the concepts of twist 
and writhe and the decomposition of the latter. 
We study the dependence of writhe on the geometrical 
properties of helical curves, paying particular attention to their
height above a plane and to the presence of dips. 
The relationship between shape, writhe, and helicity of S-shaped flux 
ropes is then discussed (Sect.~\ref{sec:shape}). In 
Sect.~\ref{sec:app} we measure the conversion of twist into writhe in several
numerical simulations of unstable magnetic flux ropes and discuss the 
implications of the results for erupting filaments and CMEs. We also point out
that estimates of the writhe of erupting filaments can easily be obtained 
from their apex rotation as soon as the apex height significantly 
exceeds the footpoint separation. Section~\ref{sec:dis} summarizes the findings.

\begin{figure}[t]
\centering
\includegraphics[width=1.0\linewidth]{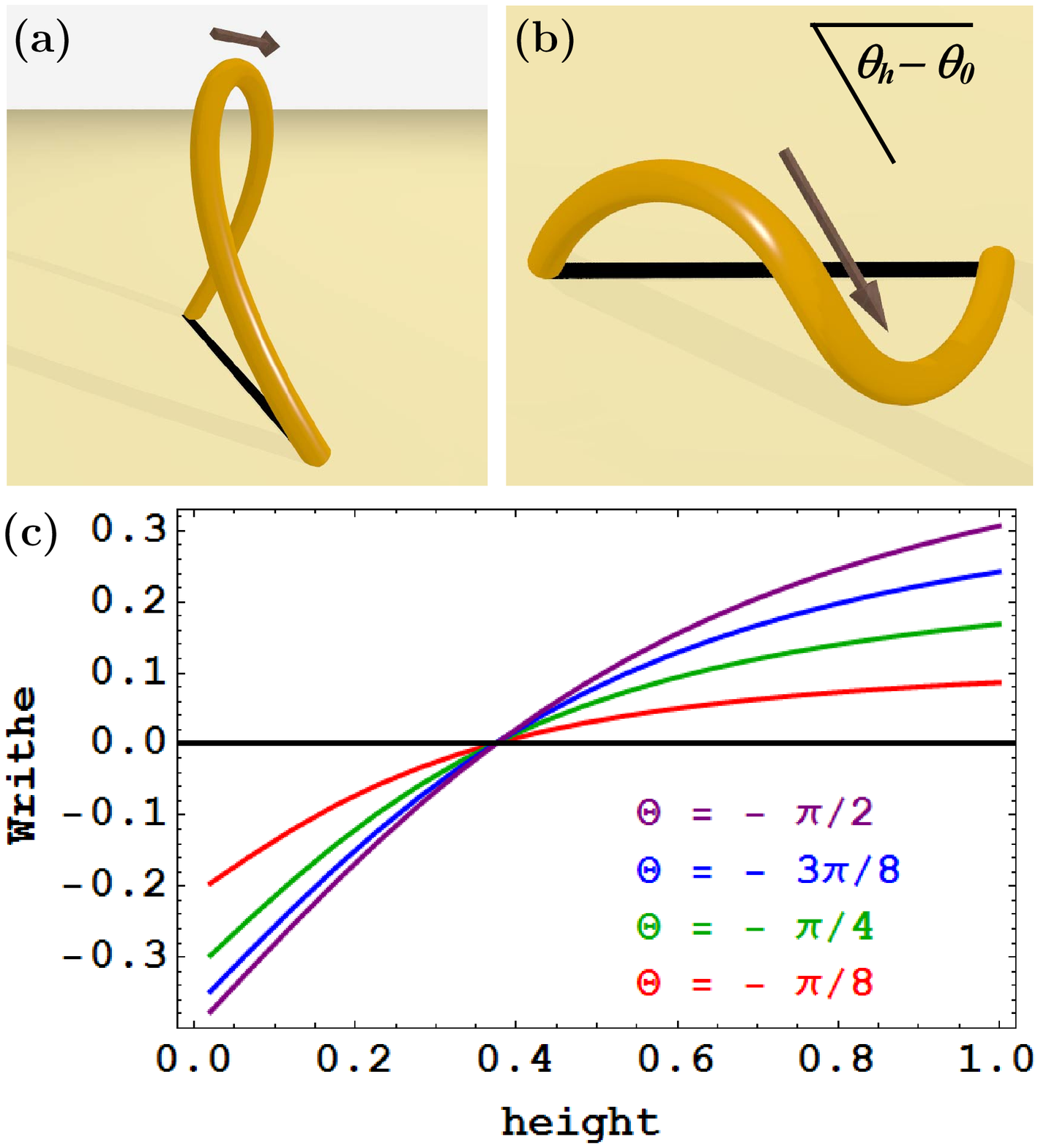}
\caption
{ 
A loop with one maximum, seen from the side {\bf (a)} and the top {\bf (b)}. 
The nonlocal writhe is proportional to the difference in angle between the 
tangent vector at the top, and the line connecting the footpoints. This angle 
is $\theta_h-\theta_0 = -\pi/3$ for the loop shown. The corresponding nonlocal 
writhe is $W_{nonlocal}= 1/3$ while the local writhes are 
$W_{1\, local} = W_{2\, local}= -0.0573$ for a total $W = 0.219$. {\bf (c):} 
The writhe for the family of curves described by 
Eqs.~(\ref{eqn:parabola1}--\ref{eqn:parabola4}). The horizontal axis gives 
the maximum height $h$ of the curves, measured in units of the footpoint
separation. For the curves shown, the nonlocal writhe 
$W_{nonlocal} = -\Theta/\pi$ is positive. All these curves display a reverse 
S shape as seen from above. 
}
\label{fig:parabolae}
\end{figure}

\section {Twist and Writhe}
\label{sec:def}

\subsection {Definitions}
\label{subsec:def}
\cite{Calugareanu:a} introduced a quantity called writhe to measure how much 
a closed curve coils and supercoils. This quantity is extensively used in DNA 
research to describe the physical configuration of the DNA molecule 
\citep{maggioni2008} (in addition to the basic double helix structure, 
the molecule must be highly coiled in order to fit into a microscopic cell). 
There are several equivalent definitions of writhe for closed curves 
\citep[e.g.][]{Aldinger95,ber06}. 
Perhaps the simplest definition to understand without getting lost in mathematical 
details involves counting crossings in pictures of a curve (see, for example, 
Fig.~\ref{fig:crossings}). If we draw a two--dimensional projection of the curve, 
it crosses itself a certain number of times. Highly coiled or tangled curves will 
show many crossings, while a non--intersecting curve confined to a plane will show 
no crossings at all (unless seen exactly edge--on). Crossings can be labelled 
positive or negative depending on the orientation of the sections of the curve 
above and below the crossing, as shown in the figure. Thus if an arrow
pointing along the direction of the lower curve rotates clockwise to match the
direction of the upper curve, then the sign is positive. Note that the sign of a
crossing stays the same if we reverse the direction of the curve, because both 
segments above and below a crossing change direction. 

Given a curve and a viewing angle, we can count the number of positive crossings 
$N_+$ and subtract the number of negative crossings $N_-$.  The result is an 
integer -- however, this integer depends on which viewing angle we are using. 
The writhe averages $N_+ - N_-$ over all viewing angles. 

As a consequence the writhe of a circular coil depends not only on how many coils 
there are, but on whether the coils are fat or thin, i.e. on the ratio between its 
height and width (Fig.~\ref{fig:coils}). For thin coils, only special viewing 
angles (from near the axis) will see crossings. But for fat coils most viewing angles 
display crossings.

While the writhe is a property of a single curve, the definition of
twist involves two curves, which can be viewed as the edges of a ribbon.
\cite{Calugareanu:a} derived a remarkable formula for the structure of a 
ribbon. Consider two closely aligned curves, like the two sides of a ribbon or 
the two strands of a DNA molecule. Assume that the ribbon or molecule closes upon
itself, so that the two curves do not have endpoints. These two curves link each 
other by some (integer) amount $L$; they also twist about each other by some 
(usually non-integer) amount $T$. Let the writhe $W$ be the writhe of one of the 
two curves (alternatively a central curve between the two). Then linking number 
equals twist plus writhe,
\begin{equation}
    L=T+W.
\end{equation}
For our purposes we need two modifications. First, we are concerned with magnetic 
flux \emph{tubes} rather than ribbons. Also, our tubes are not closed: they have 
ends (footpoints) on a boundary plane or planes. For these two reasons, we use 
modified expressions for writhe as given in \cite{ber06}. For a magnetic field, 
we can replace linking number $L$ with magnetic helicity $H$, which averages the
linking of all pairs of field lines. For a straight flux tube of flux $F$, the 
helicity is $H=TF^2$. (By definition, if field lines twist around the axis 
by an angle of $\Phi=2\pi$, then $T=1$.) The twist $T$ measures how much of the helicity 
is generated by parallel electric currents. In particular, if $s$ denotes arc 
length along the central field line of the tube, and $J_\|$ the component of 
current parallel to this field line, then $T=\int{\deriv{T}{s}\mathrm{d}s}$, with
\begin{equation}
\label{dtds}
\deriv{T}{s}=\frac{\mu_0 J_\|}{4\pi B_\|}.
\end{equation}
The writhe for a curve with endpoints (e.g. the axis of an arbitrary 
shaped flux tube) is defined by the magnetic C\u{a}lug\u{a}reanu 
formula \citep{ber84,moffatt92}
\begin{equation}
    H = \left(T + W\right)F^2.
\end{equation}
A current-free tube has zero twist, by Eq.~(\ref{dtds}). Thus we can define 
the writhe of a curve also as the helicity (divided by $F^2$) of a thin 
current-free flux tube which follows the curve.

\begin{figure*}
\sidecaption
\centering
\includegraphics[width=0.7\linewidth]{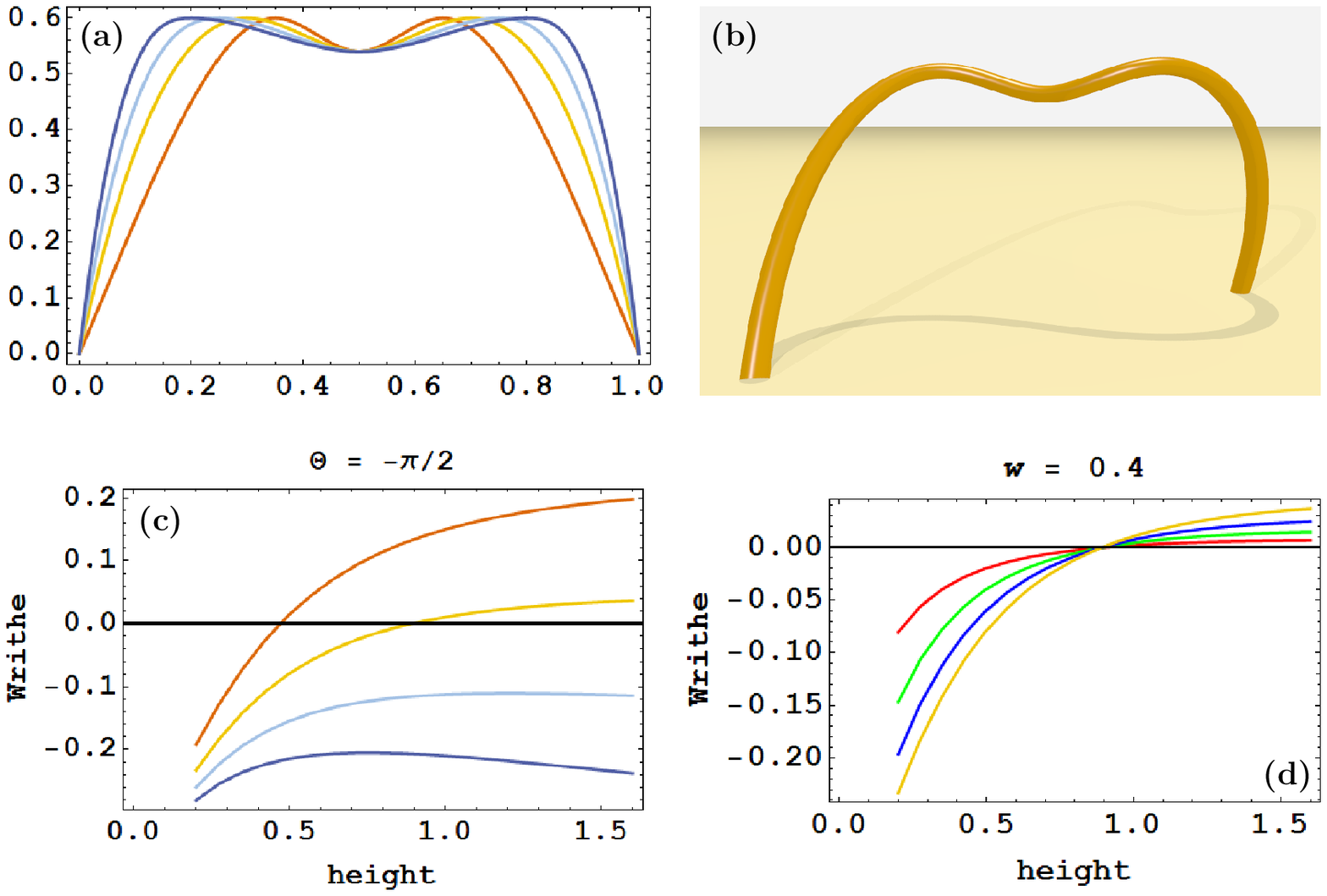}
\caption{The writhe for a family of dipped curves with height function 
given by a cubic spline (Eqs.~\ref{eqn:dip0}--\ref{eqn:dip}). All curves 
have $\mu = 0.9$.
{\bf (a):} the height function $z(s)$ with maximum height $h = 0.6$ for
$w=0.3,0.4,0.5,0.6$. 
{\bf (b):} A loop with $h=0.6$, $w = 0.4$, and $\Theta = -\pi/3$.  
The projection of this loop onto the bottom plane is the same as in 
Fig.~\ref{fig:parabolae}b, but here $W=-0.03$. 
{\bf (c):} Writhe calculated as a function of maximum height $h$ for 
fixed $\Theta = -\pi/2$. From top to bottom, the curves have 
$w=0.3,0.4,0.5,0.6$ (the colors correspond to panel [a]). 
{\bf (d):} Writhe calculated as a function of maximum height $h$ for 
fixed $w = 0.4$. From top to bottom on the left, the curves have 
$\Theta = -\pi/8$, $-\pi/4$, $-3\pi/8$, and $-\pi/2$. 
All these curves display a reverse S shape as seen from above.}
\label{fig:dip}
\end{figure*}

\subsection{Local and nonlocal writhe}
\label{subsec:l+nl_wri}
\cite{ber06} give expressions for the writhe of curves stretching between 
two parallel planes, as well as loops with both endpoints on a bottom plane. 
The method consists of separating the curve into pieces at maxima and minima 
in height $z$. Let us suppose, for simplicity, that there is just one maximum, 
at height $z=h$. We first ask how much the two pieces, or legs, rotate about 
each other while rising from the bottom to the top (see Fig.~\ref{fig:parabolae}). 
Let $\theta_h$ be the orientation of the tangent vector at the top of the loop 
with respect to the $x$ axis. Also suppose that the line from the positive 
endpoint in the bottom plane ($z=0$) to the negative endpoint has an orientation 
$\theta_0$. Then the two legs of the curve rotate about each other by 
$\theta_h - \theta_0$. The quantity 
\begin{equation}
\label{apextangent}
    W_{nonlocal} = -\frac 1 { \pi} (\theta_h - \theta_0)
\end{equation}
contributes to the total writhe. It is called the \emph{nonlocal writhe}, because 
for most of their lengths, the two legs are far away from each other. Similar 
formulae can be derived for loops with more than one maximum \citep{ber06}.

In addition, each leg on its own may contribute to the total writhe. An individual 
leg may have a helical shape, for example. The individual contributions will be 
called \emph{local writhe}. Thus for a loop with legs 1 and 2, we have the 
decomposition
\begin{equation}\label{decomp}
    W  =  W_{1\, local} + W_{2\, local} + W_{nonlocal}.
\end{equation}
Let $\tang_1(z)$ be the (unit) tangent vector to leg 1 at height $z$, 
and let $\tang '_1 = \diff \tang_1 /\diff z$ be its derivative with respect to 
height. Then one finds
\begin{equation}\label{eq:W_1,local}
W_{1\, local}  =  \frac 1 {2 \pi}\int_0^h
    \frac{1}{(1+|T_z|)}\,({\bf {T}}_1 \times {\bf {T}}'_1)_z \, 
    \diff z \label{localwrithe},
    \label{eqn:localwrithe}
\end{equation}
with a similar expression for $W_2$ \citep[see][]{ber06}.

If only the S shaped projection of the loop onto the bottom plane is 
available, as it is often the case in solar observations, we can still 
infer the sign of the local writhe. If the loop bends in a clockwise 
(counterclockwise) manner from the apex to either footpoint, the local 
writhe is positive (negative).

\subsection{Height dependence of writhe}
\label{subsec:height_wri}
To illustrate how the writhe behaves, we consider a family of basic 
loop shapes. Place the endpoints of the loops at $(x,y,z)=(-1/2,0,0)$ 
and $(x,y,z)=(1/2,0,0)$. Thus the horizontal distance between the 
endpoints is one; equivalently, all lengths will be scaled to this 
distance. Let the axis of the loop be a curve $(x(s), y(s), z(s))$ 
for $0\le s \le 1$. We choose the horizontal coordinate functions 
separately from the height function. A family of S shapes can be 
generated using the form 
\begin{eqnarray}
  x(s)  &=& (s-1/2) \cos \theta(s); \label{eqn:parabola1}\\
  y(s)  &=& (s-1/2) \sin \theta(s); \label{eqn:parabola2}\\
  \theta(s) &=& 4 \Theta s (1-s).   \label{eqn:parabola3}
\end{eqnarray}
The number $\Theta$ determines the maximum amount of rotation of the 
curve: for $\Theta = 0$ the curve remains within the $x-z$ 
plane, while for $\Theta=\pm \pi/2 $ the top of the curve is oriented 
perpendicularly to the line between the endpoints.

We also need a height function $z(s)$. Figure~\ref{fig:parabolae}a,b employs 
a parabolic shape
\begin{equation}
z(s) = 4 h s (1-s),
\label{eqn:parabola4}
\end{equation} 
where $h$ is the maximum height, and $\Theta = -\pi/3$. The curve exhibits a 
reversed S shape when seen in projection on the bottom plane 
(Fig.~\ref{fig:parabolae}b).

For curves of this form, local and nonlocal writhe give contributions of 
opposite sign. We can ask what the writhe will be as a function of height 
$h$ and maximum rotation $\Theta$ (see Fig.~\ref{fig:parabolae}c). 
For tall curves, the nonlocal writhe dominates, while for short curves 
the local writhe dominates. Thus, for a given $\Theta$, the sign of 
writhe depends on the height of the curve. For the family of curves shown 
in Fig.~\ref{fig:parabolae}, the writhe vanishes for all curves with 
$h\approx 0.37$. Hence the writhe can be zero, even if the curve exhibits 
an S shape when seen from above. 

\begin{figure}[t]
\centering
\includegraphics[width=1.0\linewidth]{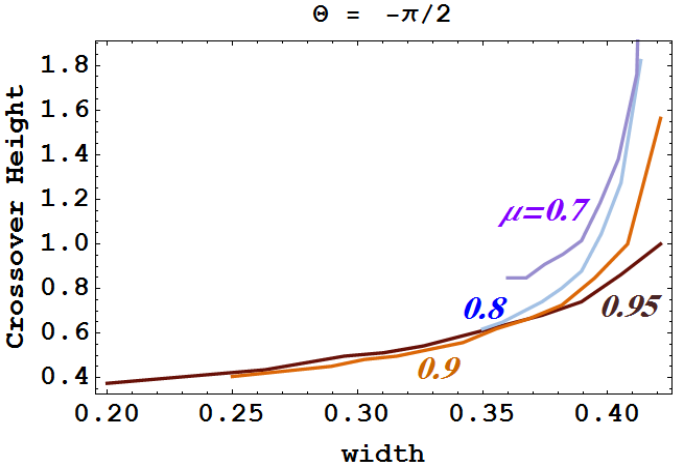}
\caption
{ 
Heights of the maxima of dipped curves at which the writhe changes sign,
for different dip strengths $\mu$, shown as a function of the distances 
between the maxima of the height function (see text for details), for
$\Theta = -\pi/2$. Note that the crossover heights are practically 
independent of $\Theta$ (Fig.~\ref{fig:dip}d). 
}
\label{fig:diprange}
\end{figure}

Changing the height function seems to make little difference, as long as 
the height function has only one maximum. 
For example, for $z(s)=h\,\mathrm{sin}(\pi s)$ 
we find almost identical values for the writhe; the crossover height where 
writhe vanishes goes down to near $h\approx 0.36$. A quadratic function 
$z(s) = h(1 - (2s-1)^4)$ gives similar results, with crossovers
 $h\approx 0.37\mbox{--}0.38$.

The writhe graphs do change if the curve has more than one maximum. Consider 
a loop with a dip at its central part. We keep the same horizontal coordinate 
functions as defined in Eqs.~(\ref{eqn:parabola1}--\ref{eqn:parabola3}),
but change the height function to a cubic spline. The spline is defined by 
the conditions
\begin{eqnarray}
	z(0) & = & z(1) = 0;\label{eqn:dip0}\\
	z(0.5-w/2) & = & z(0.5+w/2) = h;\\
	z'(0.5-w/2) & = & z'(0.5)  =  z'(0.5+w/2) = 0;\\
	z(0.5) & = & \mu h ;\label{eqn:dip}
\end{eqnarray}
in addition, second derivatives match at the extrema. The parameter $\mu$ 
gives the amount of dip, e.g. $\mu = 0.9$ gives a 10 per cent dip 
(Fig.~\ref{fig:dip}), and $w$ is the distance of the maxima. Note 
that the horizontal shape has not changed; the dip is not visible if the 
curve is seen from above (Fig.~\ref{fig:dip}b) . 

For these curves, there are two maxima and one minimum. Let the tangent 
vector orientations at the maxima be $\theta_{1\max}$ and $\theta_{2\max}$, 
with orientation at the minimum $\theta_{\min}$. Then we replace 
Eq.~(\ref{apextangent}) with
\begin{equation}\label{diptangent}
    W_{nonlocal} = -\frac 1 { \pi} (\theta_{1\max} + \theta_{2\max} 
                   - \theta_{\min} - \theta_0).
\end{equation}
For such curves, the writhe often has the same sign for all heights $h$ 
(see the bottom two curves in Fig.~\ref{fig:dip}c). Note that forward 
(reverse) S shape then gives positive (negative) writhe, for all heights. 
For some dip 
shapes (e.g., the top two curves in Fig.~\ref{fig:dip}c) there is still a 
change in sign of writhe, but at different (and usually larger) heights 
than in the curves without a dip. For example, Fig.~\ref{fig:dip}d shows 
the particular case $w=0.4$ for four different values of rotation angle 
$\Theta$. As in the curves without dips, the crossover height (here near 
$h = 0.9$) is almost completely insensitive to the rotation angle.   

Figure~\ref{fig:diprange} shows that the range of $w$, which 
yields a sign change of writhe, decreases with increasing dip strength 
$\mu$. 
For $w>0.45$, the writhe does not change sign for all $\mu\le0.95$. 
Hence, in order for a sign change of writhe at a certain
height to occur, the dip must be relatively small, both in width and depth.

\section{Sign of writhe and helicity for S-shaped coronal flux ropes}
\label{sec:shape}
The height dependence of writhe for loop-shaped curves obtained in 
Sect.~\ref{sec:def} has implications for 
the relation between S shape, sign of axis writhe, and chirality of coronal 
flux ropes, which we discuss in the following two sections. 

Since a projected S shape of a curve rooted at both ends in the photosphere
is an immediate signature of the writhing of the curve out of a plane, one 
is tempted to expect a unique relationship between the orientation of the S 
and the sign of the curve's total writhe. However, as shown in \cite{ber06} 
and in Sect.~\ref{subsec:height_wri}, this relationship is \emph{not} unique. 
The ambiguity can affect the relationship between the orientation of S-shaped 
flux ropes and their chirality, since the total axis writhe contributes to 
the helicity. This leads to the question why the solar observations indicate 
such a strong association between S shape and 
chirality.

\begin{figure*}[t]
\centering
\includegraphics[width=1.\linewidth]{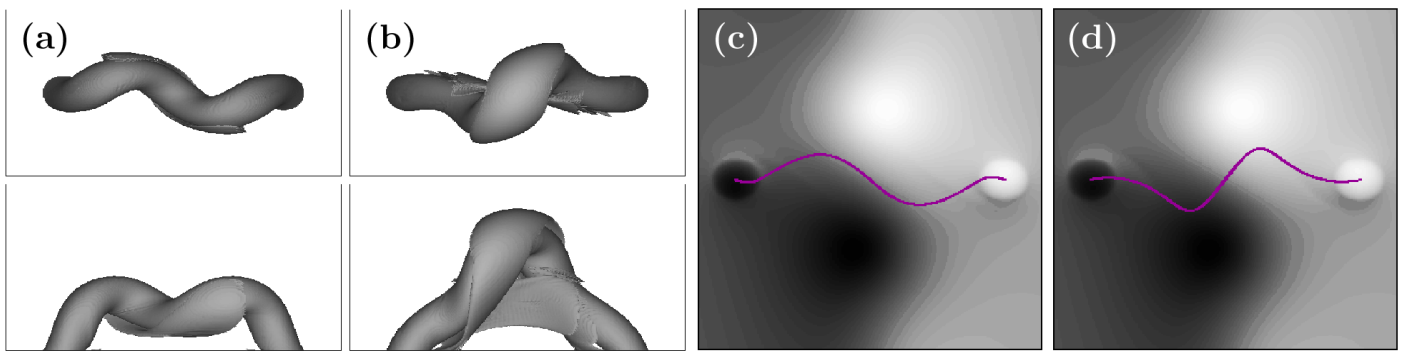}
\caption
{
{\bf (a,b):} Two kink-unstable flux ropes, displayed by isosurfaces of current 
density (from T\"or\"ok et al. 2004). Both simulations start from the 
same configuration with left-handed twist (i.e. negative helicity) and zero
axis writhe, but with oppositely directed initial perturbations. The top panels 
show a top view and the bottom panels show a side view. The rope in (a) kinks 
downward and develops a reverse S shape when seen from above. The rope in (b) 
kinks upward and develops a forward S shape when seen from above. 
{\bf (c):} Top view on the magnetic axis of the downward kinking rope, 
shown at the same time as in (a). {\bf (d)}: Same as (c) for the upward kinking 
rope. Contour plots of the vertical magnetic field in the bottom plane are
included in (c) and (d). White (black) contours show $B_z>0$ ($B_z<0$). 
}
\label{fig:kink}
\end{figure*}

Let us recast the two striking properties 
of curves with an S-shaped projection, which are apparent from the results 
in Sect.~\ref{subsec:height_wri}. First, for curves without a dip, the sign 
of the total writhe will flip if the apex height $h$ is changed in a certain 
range while the S orientation is kept (Fig.~\ref{fig:parabolae}c). 
Equivalently, if the sign of the total writhe is kept, the orientation of 
the S will flip (see Figs.~\ref{fig:kink} and \ref{fig:rot2} below). The 
flip occurs for $h$ of order 0.4 the footpoint distance, which is relevant 
for solar filaments. Second, the occurrence or disappearance of a dip (of 
rather moderate depth [Fig.~\ref{fig:dip}]) in a curve of sufficient height 
($h \gtrsim 0.4$) can flip the sign of the total writhe if the S 
orientation is kept (compare Figs.~\ref{fig:parabolae} and \ref{fig:dip}).

These results demonstrate that care is needed when deriving the sign of the
writhe from the observed orientation of S-shaped structures on the solar
disk. It must be ensured that the height of the object is estimated
correctly, and the possibility that the object's axis has a dip must be
taken into account if $h \gtrsim 0.4$.

First consider filaments. The height of stable filaments falls below the limit 
$h \approx 0.4$ in most cases. The presence or absence of a dip is then 
irrelevant for the relationship between the S orientation 
and the sign of the total writhe. For erupting filaments, on the other hand,
it is often possible to follow their evolution into the height range
$h \gtrsim 0.4$, especially in EUV observations. If filaments rise to such and 
greater heights, they hardly ever display a dip in their top part 
(see Sects.~\ref{subsec:S_erup} and \ref{subsec:wri_rot}). Thus, in practice, 
S shape and writhe can be unambiguously related to each other if a distinction 
is made between low-lying (stable) and high-arching (erupting) filaments 
and intermediate cases are excluded. 

The observed correlation between the magnetic orientation
(sinistral vs.\ dextral) of stable active-region filaments and the
orientation of their curved ends in the vicinity of sunspots with whirls
\citep{Rust&Martin1994, Zirker&al1997}
then implies that such filaments possess positive (negative) writhe
if the orientation of their axial field is sinistral (dextral). Thus,
the sign of writhe is identical to the chirality of the sunspot field
near the end of the filament. This corroborates the expectation that
such filaments are threaded by field emanating from the sunspot, which
suggests that the chirality of their field is right (left) handed for
the sinistral (dextral) orientation of the axial field. Also, it allows
for the possibility that writhe helicity contains a considerable
fraction of the helicity of the field that threads such filaments.

The above relationships also have an interesting implication for 
erupting filaments which start from 
low heights ($h \lesssim 0.4$), possess writhe (S shape) already before 
they rise and keep the sign of writhe during their evolution. 
These filaments must reverse their S orientation in the course of the rise. 
We will consider this in Sect.~\ref{subsec:S_erup}. 

The relation between S shape and chirality is likely to be simpler for
sigmoids. These sources are supposed to lie in separatrix surfaces
or quasi-separatrix layers of
the field underneath flux ropes \citep{tit99}, especially if they form a
continuous S \citep{gib04, Green&Kliem2009}. Consequently, they outline
low-lying helical field lines at the periphery of a flux rope, not the
axis of a rope. The writhe of these field lines has the same sign as
their twist (i.e., as the twist helicity of the rope), and it has a unique
relationship to the S orientation.

\section{Writhe of erupting flux ropes and filaments}
\label{sec:app}
In this section, we measure the writhe of the magnetic axis of a flux rope
in the course of ideal MHD instabilities
in numerical simulations, and we discuss the 
implications of the results for filament eruptions and CMEs. In flux rope 
geometry, the spine of a filament follows the magnetic axis of the rope 
closely \cite[e.g.,][]{aul98, Bobra&al2008}. All simulations presented below 
integrate the ideal MHD equations and use the analytical coronal flux rope 
model by \cite{tit99} as initial condition. The model consists of a line-tied, 
arched, and twisted magnetic flux tube embedded in an arcade-like 
potential field (see \citeauthor{tit99} \citeyear{tit99} for details).
The eruptions of the flux rope are driven either by the helical kink
instability (KI) alone, or by the combined action of the 
KI and the torus instability \citep[][hereafter TI]{bat78,kli06}.
Numerical diffusion permits magnetic reconnection to occur where current
layers steepen in response to the development of these ideal MHD
instabilities.
In all simulations, we measure the writhe of the rope's magnetic axis 
using Eqs.~(\ref{apextangent}--\ref{eq:W_1,local} and \ref{diptangent}).

\subsection {Kink-unstable magnetic flux ropes}
\label{subsec:kink}
Both types of curves discussed in Sect.~\ref{subsec:height_wri}, with 
and without a dip at the apex, are realised in the 3D MHD simulations 
of kink-unstable flux ropes by \cite{toe04}. Figure~\ref{fig:kink} shows 
snapshots of the current channel in the core of the flux rope
in the course of the instability for two of these simulations. 
Both simulations start from the same flux rope configuration with a 
left-handed average twist angle of $\Phi=2\pi T=-4.9\,\pi$ and vanishing
writhe of the axis (whose projection on the bottom plane is a straight
line). In the following we will use the twist angle in
quantitative statements. 
Its end-to-end value is calculated as an average over the cross-section 
of the current channel in the core of the flux rope \citep{toe04}. 

The downward kinking rope shown in Fig.~\ref{fig:kink}a has a dip in 
its middle part and displays a reverse S shape when seen from above. 
Using the parameters defined in Sect.~\ref{subsec:height_wri}, its axis can 
be described by $h=0.26$, $w=0.57$, $\mu=0.60$, and $\Theta=-\pi/4.4$. 
The rope axis runs relatively flat, has a strong dip, and the two maxima 
are relatively far away from each other. From our considerations in 
Sect.~\ref{subsec:height_wri}, we therefore expect the writhe to be 
negative. The negative sign of the writhe follows also, of course, from 
the conservation of magnetic helicity in the course of the instability. 
The axis writhe is $W=-0.26$, with $W_{local}=0.23$, $W_{nonlocal}=-0.49$ 
(the latter being made up of contributions $-0.23$ at the minimum and 
$-0.13$ at each maximum). 

The axis of the upward kinking rope develops a forward S shape when seen
from above (Fig.~\ref{fig:kink}b). It has only one maximum, at $h=0.53$. 
This height is larger than the crossover height for curves with one 
maximum (Sect.~\ref{subsec:height_wri}), hence we expect the writhe to 
be negative. Again, this follows from helicity conservation. We find 
$W=-0.30$, with $W_{local}=-0.01$ and $W_{nonlocal}=-0.29$ ($\Theta=\pi/3.5$). 
Note that $W_{local}$ and $W_{nonlocal}$ have the same sign, which is never 
the case for the family of S-shaped curves with one maximum discussed in 
Sect.~\ref{subsec:height_wri}. 
This discrepancy is due to the fact that the flux rope axis in the simulation 
is not perfectly S-shaped when seen from above (Fig.~\ref{fig:kink}c,d).
Rather, close to the footpoints, the axis bends in the direction opposite 
to the overall orientation of the S, contributing negative local writhe.
This additional bending is an indication 
for the occurrence of KI eigenmodes with axial wavenumbers $k>1$ 
\citep[see, e.g.,][]{lin98}, which seems plausible given the relatively 
large twist used in the simulations. As a consequence, the local writhe 
changes sign as one follows the axis from the apex to either footpoint, 
which is not the case for the perfectly S-shaped curves discussed in
Sect.~\ref{subsec:height_wri}.

Note that the writhe of the two kinked ropes is very similar. 
Thus, the two simulations illustrate that flux ropes of equal
writhe can have opposite orientations of their projected S-shape, 
depending on the apex height. 

Let us now consider the amount of twist that is converted into writhe in
the course of the KI for the cases shown in Fig.~\ref{fig:kink}. Since 
the KI is an ideal MHD instability, both the axial flux and the helicity 
of the flux rope must be conserved, i.e., twist must be converted into an 
equal amount of writhe of the same sign. However, we must take care when 
calculating the twist converted within the entire rope, since flux 
surfaces far away from the axis might bend in a different way than 
the axis (for example if the flux rope expands significantly). In the 
simulations shown in Fig.~\ref{fig:kink}, this is not the case, so we can 
estimate the twist converted within the entire rope (not just in the vicinity 
of its axis) from the axis writhe. The writhe $W=-0.26$ ($W=-0.30$) 
for the downward (upward) kinking rope corresponds to a converted twist of 
$\Phi=-0.52\,\pi$ ($\Phi=-0.60\,\pi$). In order to check the reliability of 
this estimate, we independently measure the flux rope twist for the upward 
kinking case at the stage of the evolution shown in Fig.~\ref{fig:kink}b, 
using Eqs.~(12--13) in \cite{ber06}. We find that the converted twist 
measured that way and the value estimated from the axis writhe differ by 
less than 1\%, as long as we exclude flux surfaces very close to the
rope surface, where the twist is strongly nonuniform \citep[see Fig.~2
in][]{toe04}. This shows that the 
twist converted in the course of the KI can be reliably estimated from the 
axis writhe if the inhomogeneity of the radial twist profile and its 
changes remain modest. It also shows that the helicity is conserved to a 
high degree of accuracy in the simulation. 

Interestingly, the amount of converted twist of $\approx (0.5-0.6)~\pi$ at
the stage of the instabilities shown in Fig.~\ref{fig:kink} is smaller (only 
$\approx 10~\%$ of the initial twist) than one might intuitively expect
from the relatively strong deformation of the flux ropes. It is often
assumed that the KI converts a twist of $\pm 2~\pi$, 
which probably arises from the fact that helically deformed structures on 
the Sun typically exhibit one helical turn, and are therefore often believed 
to have a writhe of $\pm 1$ \citep[e.g.][see also Sect.~4.2]{rus03}. However, 
as Fig.~\ref{fig:coils} shows, the number of helical turns of a curve or flux 
rope can be very different from its writhe, so that estimations of writhe 
(and of twist converted in the course of the KI) cannot be made from the 
observed number of turns alone. In the following subsection, we will investigate 
the conversion of twist into writhe in the course of flux rope 
instabilities in more detail.

\begin{figure}[t]
\centering
\includegraphics[width=1.0\linewidth]{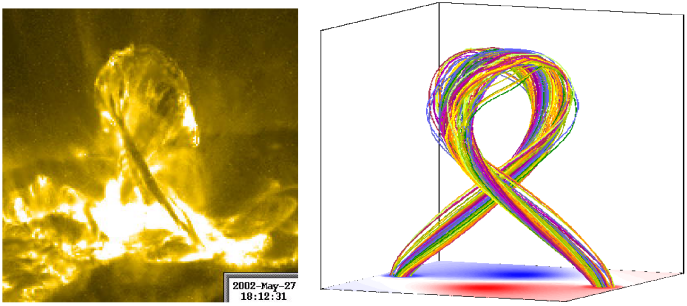}
\caption
{
Confined filament eruption on 2002 May 27 observed by {\em TRACE} in the 
195\,{\AA} band,
and magnetic field lines outlining the core of the kink-unstable flux rope 
in the confined eruption simulation by T\"or\"ok \& Kliem (2005). The flux 
rope axis in the simulation has a writhe of $W=0.67$ at the state shown 
($t=37\tau_\mathrm{A}$ [Alfv\'en times], apex height = 1.1 times 
the footpoint distance; see also Fig.~\ref{fig:tk05}a). Note that the
flux rope twist in the simulation was chosen right-handed, in order to
account for the morphology of the kinked filament. 
}
\label{fig:event1}
\end{figure}
\begin{figure}[t]
\centering
\includegraphics[width=1.0\linewidth]{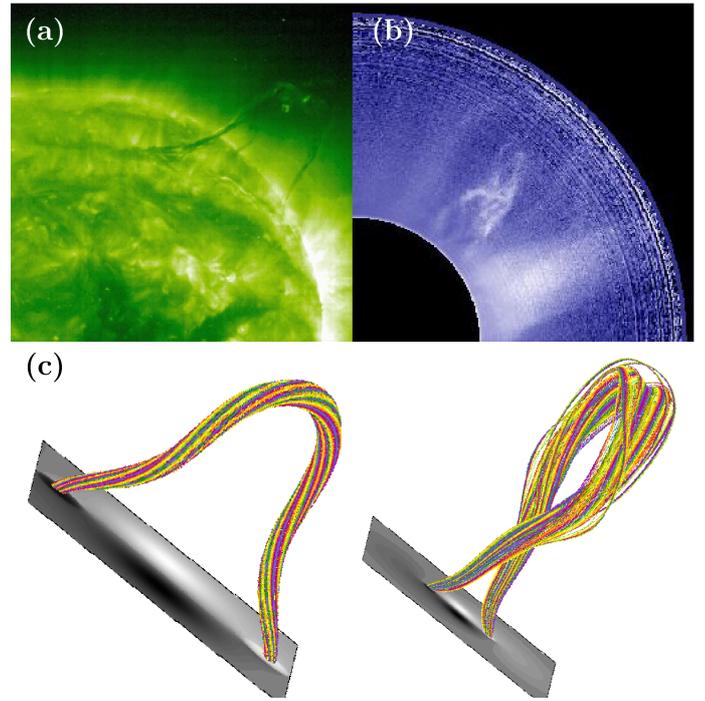}
\caption
{
{\bf (a):} Filament eruption on 2003 February 18 observed by the EUV 
Imaging Telescope (EIT) onboard the \textsl{Solar and Heliospheric Observatory} 
in the 195~{\AA} band at 02:12~UT.
{\bf (b):} The same filament as in (a), observed as CME core by HAO's 
Mauna Loa Solar Observatory Mk4 white-light coronagraph at 02:34~UT. 
The filament as seen by EIT does not show indications for the occurrence 
of a significant rotation about its rise direction, whereas the CME 
core does (note the crossings of its legs). 
{\bf (c):} Two snapshots of the CME simulation in 
T\"or\"ok \& Kliem (2005) at the same viewing angle but 
different spatial scale, taken at $t=22\tau_\mathrm{A}$ (left) and
$t=43\tau_\mathrm{A}$ (right).
The apex height of the flux rope axis is, respectively, 0.88 and 3.8
of the footpoint distance (which is $\approx\!3.3$ times the initial
apex height). The total (nonlocal) writhe of the 
axis is, respectively, $-0.254$ ($-0.212$) and $-0.623$ ($-0.518$). 
}
\label{fig:event2}
\end{figure}

\subsection {Writhe in confined and ejective eruptions}
\label{subsec:erup}
\cite{toe05} presented two simulations of erupting kink-unstable flux 
ropes with identical initial average flux rope twist, $|\Phi| \approx 5\pi$, 
and nearly identical geometrical rope parameters, but 
different ambient potential fields. In the first simulation (hereafter 
TK1; Fig.~\ref{fig:event1}), the potential field decays only slowly with 
height above the initial flux rope position. The rope ascends driven by 
the KI (similar to the case shown in Fig.~\ref{fig:kink}b), but its rise 
terminates at an apex height of $\approx 3.5$ times the initial height 
$h_0$, in very good quantitative agreement with a confined filament eruption 
observed on 2002 May 27 \citep[see][]{ji03}. 

In the second simulation (hereafter TK2; Fig.~\ref{fig:event2}), the 
potential field decreases much faster with height. As a result, the TI 
sets in once the KI has lifted the rope to a height where the 
potential field drops off sufficiently steeply. The rope is then additionally 
accelerated by the TI and eventually ejected. The flux rope rise 
characteristics could be scaled to closely match the acceleration profile of 
a CME on 2001 May 15 \citep[see][]{mar04}. 

\begin{figure}[t]
\centering
\includegraphics[width=1.00\linewidth]{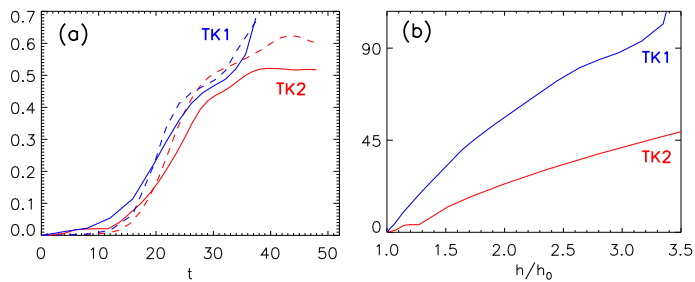}
\caption
{
{\bf (a):} Absolute values of total writhe (dashed) and nonlocal writhe 
(solid) as a function of time for the confined (blue) and ejective (red) 
flux rope eruption simulations in T\"or\"ok and Kliem (2005). 
{\bf (b):} Absolute values of flux rope axis apex rotation in degrees 
for the two simulations, as a function of axis apex height normalized to 
its initial value. Note that the rotation is proportional to the nonlocal 
writhe (a nonlocal writhe of $\pm 0.5$ corresponds to a rotation of $\mp 
90^\circ$). We plot absolute values here since different signs (or handedness) 
of the initial flux rope twist were used in the simulations.
Time is given in Alfv\'en times based on the initial apex height of
the flux rope axis and the initial Alfv\'en velocity at this point,
$\tau_\mathrm{A}=h_0/V_\mathrm{A0}$.
}
\label{fig:tk05}
\end{figure}

Here we measure the evolution of the writhe of the flux rope axis in these 
simulations. As in the simulations described in Sect.~\ref{subsec:kink},
the writhe vanishes initially. Figure~\ref{fig:tk05} shows the evolution of 
$W$, of $W_{nonlocal}$, and of the apex rotation. The rotation is proportional 
to $W_{nonlocal}$, since the axis has only one maximum at all times. The 
writhe first grows exponentially, followed by a transition to a saturation phase. 
The second strong increase in TK1 at $t\approx35\tau_\mathrm{A}$ is caused by an
additional deformation of the flux rope axis due to the onset of magnetic 
reconnection with the overlying field around this time \citep[see][]{toe05}.
$W_{nonlocal}$ reaches $\approx 0.5$ in TK2, corresponding to a rotation of 
$\approx 90^\circ$. A very similar value is found for the rope in TK1 before 
it starts to reconnect with the overlying field. 
As expected from Sect.~\ref{subsec:height_wri}, the nonlocal 
writhe clearly dominates the local writhe once the flux rope has sufficiently 
risen. In both simulations, the rope develops a clear S shape when viewed in 
projection on the bottom plane. 

The temporal evolution of the writhe is very similar in TK1 and TK2, since 
the growth rate of the KI, largely set by the initial twist, is nearly the 
same (Fig.~\ref{fig:tk05}a). The evolution 
as a function of height, however, is quite different (Fig.~\ref{fig:tk05}b). 
The additional acceleration of the rope by the TI spreads the rotation of 
the apex over a larger height range \citep[compare also Figs.~1 and 4 in]
[]{toe05}. The TI, which is a form of the lateral kink instability,
primarily expands the unstable flux loop, while the KI primarily
produces a helical shape.

Given the nearly identical choice of the initial flux rope parameters 
in the two simulations, one can conclude from Fig.~\ref{fig:tk05}b that
a stronger field immediately above the initial rope does not only resist
the evolution into a CME more efficiently, but also produces a more
pronounced writhing at low heights. This association is opposite to the 
suggestion in \cite{Sturrock&al2001} and \cite{fan05} that the writhing 
facilitates the rupture of an unstable flux rope through the overlying 
field (which is of course strongest at low heights), and it underlines 
the importance of magnetic reconnection below the rope in permitting the 
erupting flux rope to pass through the overlying field 
\cite[e.g.,][]{Lin&Forbes2000, Vrsnak2008}. 

Since the writhing in CMEs tends to be distributed over a large height range 
(Fig.~\ref{fig:tk05}b), much of it may escape detection. 
Typically, the writhing is apparent from the apex rotation of an 
associated erupting filament or prominence observed in H$\alpha$ or in the 
EUV, and this is usually limited to the low and middle corona. Therefore, 
a significant rotation (and the KI) may occur in a larger fraction of CMEs 
than usually thought. 

An illustrative example, the filament eruption and CME on 2003 February 18 
\citep{fan05, rus05}, is shown in Fig.~\ref{fig:event2}. The EIT data do not 
yield indications of a significant rotation, but the Mk4 coronagraph reveals 
that the legs of the CME core cross to form an ``inverse $\gamma$'' shape 
when the core reaches a height of 
$\approx 1 R_\odot$ above the solar surface. Such a shape develops due to
writhing and is commonly regarded to be evidence of the KI
\citep[e.g.,][]{gil07}.
  
We note that the inverse $\gamma$ shape has been associated for the two 
events shown in Figs.~\ref{fig:event1} and \ref{fig:event2} with a writhe 
of $+1$ and $-1$, respectively \citep{rus03,rus05}. However, the values 
obtained from the simulations indicate that the writhe of such structures 
is rather some number between 0 and 1. This appears plausible if one 
recalls that the writhe can be expressed as average crossing number (see 
Sect.~\ref{subsec:def} and Fig.~\ref{fig:crossings}). Consider, for example, 
the filament shown in Fig.~\ref{fig:event1}: from the particular viewing 
angle of the observation we see one (positive) crossing of the axis. 
Other viewing angles will display either one crossing or 
no crossing, so that the average crossing number, i.e. the writhe, 
must be less than unity.

\subsection{Changing S shape}
\label{subsec:S_erup}
The simulations presented in Sects.~\ref{subsec:kink} and \ref{subsec:erup} 
all start from a flux rope which is straight in projection on the bottom plane.
Filaments, however, often exhibit an S shape already in their equilibrium 
state. In general, the spine of such filaments is a curve of nonvanishing writhe. 
In considering their eruption, we assume that the magnetic field has formed 
a flux rope topology before or in the early stages of the eruption. 
If the filament erupts from a low height ($h \lesssim 0.4$), keeping the 
sign of writhe (which is expected if one chirality dominates the field in the 
source volume), then, according to Sects.~\ref{sec:def} and \ref{sec:shape}, 
a change of the S orientation must occur. 

\begin{figure}[t]
\centering
\includegraphics[width=1.00\linewidth]{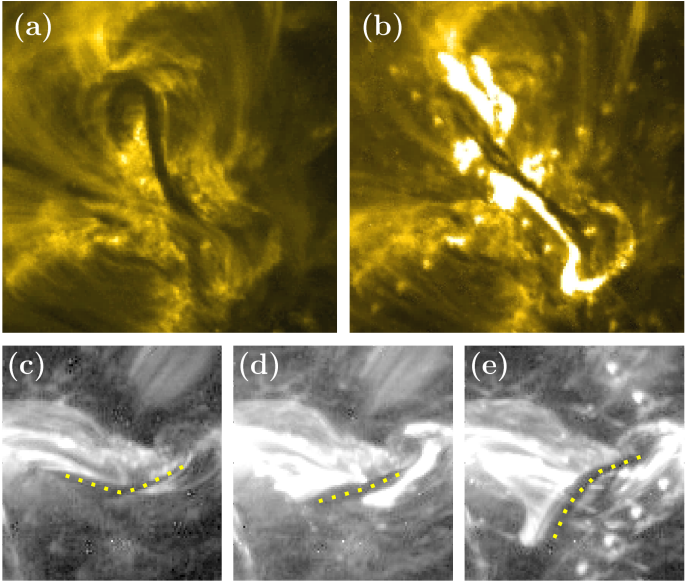}
\caption
{
Top row: Filament eruption on 2001 June 15 observed by {\em TRACE} in 195\,\AA.
{\bf (a):} At 09:57 UT, just before the eruption. The filament exhibits
a clear inverse S shape in its upper part. 
{\bf (b):} At 10:06, during the eruption. The shape of the filament has 
straightened out. Flare ribbons have formed.
Bottom row, {\bf(c--e)}: Filament eruption on 2000 June 6 observed by 
{\em TRACE} in 171\,{\AA} at 13:29:27, 13:31:34, and 13:32:20~UT, respectively. 
The filament (outlined by yellow dots)
changes its orientation from a reverse J to a forward J. 
}
\label{fig:event3}
\end{figure}

Figure~\ref{fig:event3}a--b shows a filament eruption in which an initial 
strong reverse-S shaped bending is completely straightened out. A change 
to a forward S may have followed, suggested by the rapid rotation of
the filament. However, the eruption occurred about 700~arcsec from disc
center, so that the expected transition to a forward S could not be witnessed. 
Figure~\ref{fig:event3}c--e shows an erupting filament that formed 
an incomplete S. As expected, it changed its J shape from reverse to forward 
while rising. See \citet{rom05}, \citet{rus05}, and \citet{gre07} for 
descriptions and discussions of the apex rotation and the magnetic environment 
in these events.

\begin{figure}[t]
\centering
\includegraphics[width=1.0\linewidth]{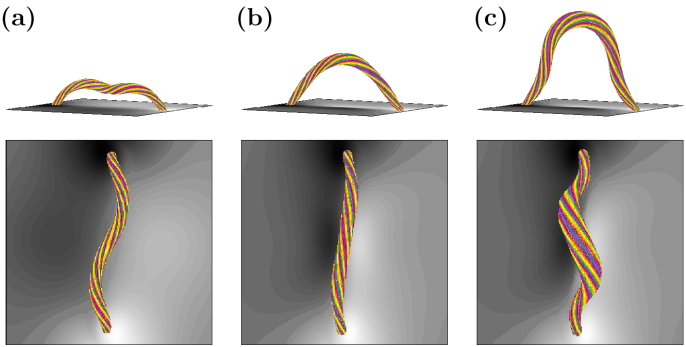}
\caption
{
Perspective and top view on magnetic field lines outlining the flux rope core 
in the simulation described in Sect.~\ref{subsec:S_erup}. 
{\bf (a):} after the initial relaxation (time reset to $t=0$). 
The two maxima of the flux rope axis have a height of 
$0.17$ of the footpoint separation and $\Theta=-\pi/3.5$.
{\bf (b)} and {\bf (c):} in the course of the eruption, at $t=55\tau_\mathrm{A}$ 
and $64\tau_\mathrm{A}$, respectively. 
The writhe is $W=-0.07$, $0.01$, and $-0.15$ (from left to right). 
}
\label{fig:rot2}
\end{figure}

Observationally, it is difficult to confirm the dynamic transition from one 
complete S shape to the other \cite[e.g.,][]{gre07}. In order to demonstrate 
the transition, we perform a 
simulation similar to the ones described in Sect.~\ref{subsec:erup}, but now 
starting from an initially S-shaped flux rope. To this end, we construct a stable 
numerical equilibrium of a modified version of the Titov \& D\'emoulin (T\&D) 
model, which contains such a rope. The rope has left-handed average twist of 
$\approx 4.7\,\pi$, similar to the initial twist in the simulations described 
above. Note that the helical shape of the constructed flux rope permits a 
larger twist in stable equilibrium than the original, toroidal T\&D rope of 
identical aspect ratio in rather similar ambient field.
We trigger the eruption of the flux rope by imposing slow, quasi-static 
converging flows towards the polarity inversion line in the bottom boundary.
The construction of the modified equilibrium and the details of the simulation
will be described in a separate publication. Here we are merely interested in 
the evolution of writhe and S shape of the erupting flux rope.

Initially, the flux rope has a dip in its middle part and displays a clear
reverse S shape when seen from above (Fig.~\ref{fig:rot2}a). Using the 
parameters discussed in Sect.~\ref{subsec:height_wri}, its axis can be 
characterised by $\Theta=-\pi/3.5$, $h=0.17$, $\mu=0.91$, and $w=0.43$. 
Inspecting Fig.~\ref{fig:dip}, we expect an axis writhe of about $-0.2$
for these values. Measuring the axis writhe, we find $W=-0.07$.
As in the above simulations, the writhe is smaller than expected from the 
curves discussed in Sect.~\ref{subsec:height_wri}, since also here the axis 
bends opposite to the overall S close to the flux rope footpoints. However, 
only the sign of the writhe is important for the present discussion. 

As the converging flows are applied, the flux rope first rises slowly 
in response to the progressive weakening of the tension of the overlying 
potential field. After $t\approx45\tau_\mathrm{A}$ it accelerates rapidly, 
most likely due to the onset of the KI or TI (we refrain here from a 
detailed investigation of the acceleration mechanism). As the rope rises, 
its shape first straigthens out, similar to the filament eruptions shown 
in Fig.~\ref{fig:event3}, and the axis writhe goes to zero. At the state 
shown in Fig.~\ref{fig:rot2}b ($t=55\tau_\mathrm{A}$), the axis has developed 
a single maximum (i.e., the dip has disappeared), its height is $0.27$, and 
its writhe is $W=0.01$. Subsequently, the rope rapidly develops a forward 
S shape, and the writhe becomes negative again. At the state shown in 
Fig.~\ref{fig:rot2}c ($t=64\tau_\mathrm{A}$), one finds $h=0.69$ and $W=-0.15$. 
Except for the short period when the writhe is close to zero (as is expected 
since the axis becomes almost straight in projection in the reversal), it stays 
always negative during the rise of the rope. Therefore, a transition from a 
reverse to a forward S shape must occur. 

\begin{figure*}[t]
\sidecaption
\centering
\includegraphics[width=1.0\linewidth]{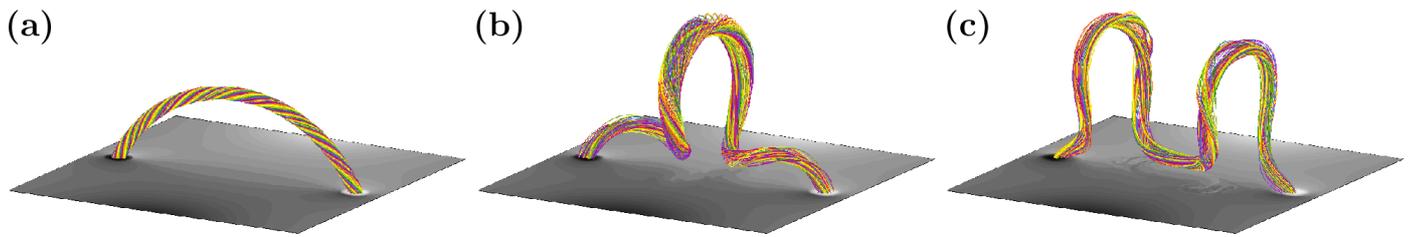}
\caption
{
Snapshots of two simulations of kink-unstable flux ropes with an initial average 
twist of $\Phi=-11\pi$. Magnetic field lines outlining the flux rope core as well 
as contours of the normal magnetic field in the bottom plane are shown. 
{\bf (a):} Initial configuration.
{\bf (b):} Upward kinking flux rope.
{\bf (c):} Downward kinking flux rope.
}
\label{fig:two_turns}
\end{figure*}

A qualitatively similar transition occurred in the simulation of a kink-unstable 
erupting flux rope by \citeauthor{fan05} (\citeyear{fan05}; see her Fig.~6a--b), 
although the flux rope equilibrium, the external potential field, and the 
driving in the photospheric boundary all differed from our simulation. Note that 
Fan displays only field lines below the magnetic axis of the flux rope, but 
these extend up to the vicinity of the axis, so that the actual rope axis must 
behave in the same manner. Also note that the straightening of the S occurs 
in the slow-rise phase for both simulations. These results have interesting 
consequences for the occurrence of the KI in filament eruptions. 

First, it is important to note that both systems are  constructed such that 
only one sign of helicity is present at the onset of the rise. Thus, the initial 
reduction of the writhe (from a negative value to nearly zero) cannot be 
driven by a conversion of flux rope twist into writhe under conservation of 
helicity, since the twist has the same sign as the initial writhe. In other words, 
if an equilibrium flux rope possesses writhe (i.e., an S shape), its initial slow 
rise, which must reduce the S shape to a nearly straight rope, cannot be caused by 
a helical kink instability, since this instability transforms twist into writhe. 
A different process is required, which actually transforms writhe into twist. 
The details of this process will likely require further study, but it is obvious 
that the converging flows in the simulation of Fig.~\ref{fig:rot2} and the 
continuing emergence of a twisted flux rope through the bottom boundary in 
\cite{fan05} were the drivers in the two simulations. 

Second, the ability of the S-shaped flux rope to remain in stable equilibrium 
for a higher twist than a straight flux rope could accommodate, and the writhe 
conversion into twist in the slow-rise phase, both support the occurrence of 
the KI after the flux rope has become straight. The two simulations demonstrate 
a scenario that enables the KI in a system which is driven significantly beyond 
the threshold of the instability by the straightening in the slow-rise phase, 
permitting a significant growth rate.

\subsection{Writhe estimate for rotating eruptive filaments} 
\label{subsec:wri_rot}
The dominance of the nonlocal writhe over the local writhe for tall 
curves discussed in Sect.~\ref{subsec:height_wri} 
provides the possibility of estimating the writhe of rotating 
filaments that erupt near Sun center even if no 3D observations are 
available. When the filament has reached a sufficient height, the writhe 
can be approximated by the nonlocal writhe (see Fig.~\ref{fig:tk05}a). 
The latter can easily be obtained from Eq.~(\ref{apextangent}) by inserting 
the observed rotation angle (Fig.~\ref{fig:parabolae}b). For this 
estimate, one has to verify that the erupting filament does not possess 
a dip in its upper part in the relevant height range. 

Limb observations of erupting filaments and prominences do not indicate 
the presence of such a dip. We can ask under which circumstances a central
dip in an erupting flux rope might occur. In the simulation described in 
Sect.~\ref{subsec:S_erup}, the flux rope has a dip prior to eruption. 
As the rope rises, the dip disappears early in the evolution, long before 
the rope reaches a height where the nonlocal writhe dominates the local 
writhe (Fig.~\ref{fig:rot2}b). A dip in the central part of a flux rope 
will naturally develop if the rope is kink-unstable and is perturbed such 
that its apex moves downward, as in the simulation shown in Fig.~\ref{fig:kink}a. 
However, in this simulation the rope does not erupt, rather its central part 
is continuously moving downward while the remaining parts of the rope do 
not ascend significantly.

It is well known that a kink-unstable flux rope can develop more than one 
helical turn if it is sufficiently twisted \citep[e.g.][]{lin98}. For a 
flux rope with two turns, a dip might form between the two helical turns. 
To check this, we perform a simulation similar to the ones described in
Sects.~\ref{subsec:kink} and \ref{subsec:erup}, but now we consider a very 
large initial flux rope twist of $\Phi=-11\pi$ (Fig.~\ref{fig:two_turns}a). 
As expected, the rope undergoes a strong helical deformation. 
However, no dip forms in its central part. 
Rather, two dips develop along the legs of the rope, while its central 
part ascends (Fig.~\ref{fig:two_turns}b). For such a configuration, care 
must be taken when estimating the writhe from the apex rotation, since the 
axis rotation at the other extrema contributes to the nonlocal writhe as
well. However, such a strong writhing is only very rarely observed 
(see, e.g., \citeauthor{rom03} \citeyear{rom03} for a filament
eruption that corresponds quite well to the simulation in
Fig.~\ref{fig:two_turns}b).

When the simulation is repeated with a small, downward directed initial
velocity perturbation, a remarkably different shape is obtained. 
Now the rope develops a large dip in its 
central part. As in the simulation shown in Fig.~\ref{fig:kink}a, the 
central section moves downward. However, in contrast to that simulation, 
the remaining parts of the rope now rise significantly 
(Fig.~\ref{fig:two_turns}c). Still, they do not succeed in pulling the 
central part upward, that part rather keeps moving downward until it hits 
the bottom plane---a situation never observed on the Sun.

One can construct a rising flux rope with dipped apex by making it 
simultaneously \emph{strongly} unstable with respect to the KI and TI. 
While such a situation can be realized in laboratory and corresponding 
numerical simulations of filament eruptions \citep{Bellan&Hansen1998, 
Arnold&al2008}, it appears very special and, hence, unlikely to occur 
under solar conditions. 

We conclude that the writhe of erupting filaments can be reliably 
estimated from the apex rotation in the relevant height range where the 
nonlocal writhe dominates, unless the filament develops a very 
particular shape with multiple maxima or minima.

\section {Summary} 
\label{sec:dis}
The availability of 3D observations provided by the \textsl{STEREO} spacecraft, 
combined with recently developed analytical expressions, facilitates obtaining 
the writhe of helical structures in the solar corona. 
In order to explore the relevance of this quantity for coronal phenomena, 
we investigated how it relates to the projected shape of helical curves 
and measured its evolution in numerical simulations of ideal MHD flux rope 
instabilities. Our results and their implications for stable and eruptive 
coronal objects can be summarized as follows.

(1) The relation between writhe and projected S shape of a curve with both 
end points in a plane is \emph{not} unique. It 
depends on the height of the curve and on the presence or absence of dips. 
Therefore, in principle, care must be taken when associating a sign of writhe 
to an observed S shape on the Sun. However, we demonstrated that this ambiguity 
does not affect low-lying filaments, as long as their height remains
below about 0.4 their footpoint distance, or soft X-ray sigmoids. This
supports the established empirical rule which associates stable forward
(reverse) S shaped structures low in the corona with positive (negative)
helicity. 

(2) Kink-unstable erupting flux ropes transform a far smaller fraction of 
their twist helicity into writhe helicity than often assumed. The writhe 
and the number of turns of a helical object differ in general. In 
particular, a simple leg crossing of a rising filament in sky projection 
can not be taken as evidence that the writhe of the filament is close to 
unity.

(3) Confined flux rope eruptions tend to show stronger writhe at low heights 
than ejective eruptions (CMEs), which acquire writhe over a larger height 
range. This argues against suggestions that the writhing facilitates the 
rise of unstable flux ropes through the overlying field, and it implies 
that a significant rotation may occur in a larger fraction of CMEs than 
suggested by the apex rotation of associated erupting filaments low in 
the corona.

(4) Erupting filaments which are S shaped already before the eruption and 
keep the sign of their axis writhe (which is expected if field of one 
chirality dominates the source volume of the eruption), must reverse their 
S shape in the course of the rise. In flux rope topology, the initial 
straightening up to the reversal represents a conversion of writhe into 
twist, which cannot be caused by the helical kink instability. However, 
the writhe conversion can be a mechanism that triggers the instability 
after the straightening.

(5) The writhe of rising loops of simple shape (which do not have 
secondary maxima or minima) can be estimated from the angle of 
rotation about the direction of ascent, once the apex height exceeds the 
footpoint separation significantly. This provides a convenient means to 
estimate the writhe of erupting filaments which rise towards the observer, 
even if no 3D observations are available.

We emphasize that for flux tubes where the twist varies significantly 
with radius, the interplay between twist, writhe, and helicity becomes 
more difficult \citep{longcope2008}. Usually an average of the 
twist over the cross section of a flux tube is of interest, while the 
writhe is a property of a single field line, in this context the 
reference field line of the twist.

This work shows that writhe is a useful quantity in interpreting S 
shaped coronal structures and in constraining eruption models. It can 
straightforwardly be computed for numerical data and can often be 
estimated from observations. It has a range of further 
applications in solar physics, for example, the study of the dynamo 
\citep{Asgari-Targhi&Berger2009}, of flux ropes rising through the 
convection zone \cite[e.g.,][]{lin98}, and of magnetic loops 
connecting different active regions, including transequatorial loops 
\cite[e.g.,][]{Chen&al2007}.

\begin{acknowledgements}
We thank the referee for his/her useful suggestions and L. van Driel-Gesztelyi 
and P. D\'emoulin for many helpful discussions.
TT and BK acknowledge an invitation to the University of New Hampshire and 
helpful discussions with T.~G.~Forbes.
This work was supported by an STFC Rolling Grant, the DFG,
and NASA grants NNH06AD58I and NNX08AG44G. 
Financial support by the European Comission through the SOLAIRE network 
(MTRM-CT-2006-035484) is gratefully acknowledged.
The research leading to these results has received funding from the
European Commission's Seventh Framework Programme (FP7/2007-2013) under
the grant agreement n 218816 (SOTERIA project, www.soteria-space.eu).
\end{acknowledgements}

\bibliographystyle{aa}
\bibliography{toeroek1f}  

\end{document}